\documentstyle[twocolumn,seceq,bm,graphicx,xspace]{jpsj}

\title{
Antiferromagnetic Order in Disorder-Induced\\ Insulating Phase of SrRu$_{\bm{1-x}}$Mn$_{\bm{x}}$O$_{\bm{3}}$ (${\bm{ 0.4\le x \le 0.6}}$)
}

\author{
Makoto \textsc{Yokoyama},$^{1}$\footnote{E-mail address: makotti@mx.ibaraki.ac.jp} Chimato \textsc{Satoh},$^{1}$ Akinori \textsc{Saitou},$^{1}$ Hirofumi \textsc{Kawanaka},$^{2}$\\ Hiroshi \textsc{Bando},$^{2}$ Kenji \textsc{Ohoyama}$^{3}$ and Yoshikazu \textsc{Nishihara}$^{1}$
}
\def\runtitle{}
\def\runauthor{Makoto {\sc Yokoyama} {\it et al}.}

\inst{
$^{1}$Faculty of Science, Ibaraki University, Mito 310-8512\\
$^{2}$National Institute of Advanced Industrial Science and Technology, Tsukuba, Ibaraki 305-8568\\
$^{3}$Institute for Materials Research, Tohoku University, Sendai 980-8577\\
}

\abst{
We have performed powder neutron diffraction measurements on the solid solutions of SrRu$_{1-x}$Mn$_x$O$_3$, and found that the itinerant ferromagnetic order observed in pure SrRuO$_3$ changes into the C-type antiferromagnetic (AF) order with nearly localized d electrons in the intermediate Mn concentration range between $x=0.4$ and 0.6. With increasing $x$, the AF moment is strongly enhanced from 1.1 $\mu_{\rm B}$ ($x=0.4$) to 2.6 $\mu_{\rm B}$ ($x=0.6$), which is accompanied by the elongation of the tetragonal $c/a$ ratio. These results suggest that the substitution of Mn for Ru suppresses the itinerant characteristic of the d electrons, and induces the superexchange interaction through the compression in the $c$ plane. We have also found that the magnetic and transport properties observed in our tetragonal samples are similar to those of recently reported orthorhombic ones.
}

\recdate{\today}
\kword{metal-insulator transition, SrRuO$_3$, ferromagnetism, antiferromagnetism, neutron scattering}

\begin{document}
\maketitle
Strontium ruthenate shows a wide variety of physical properties attributed to the itinerant feature of Ru 4d electrons and the strong mixing between Ru 4d and O 2p electrons.\cite{rf:Maeno94,rf:Mackenzie2003,rf:Grigera2001,rf:Cao97-1,rf:Callaghan66,rf:Kanbayasi76,rf:Allen96} The distorted perovskite compound SrRuO$_3$ is known to be a ferromagnet with the Curie temperature of $\sim 160\ {\rm K}$ and the ordered moment of $ \sim 1.1\ \mu_{\rm B}$.\cite{rf:Callaghan66,rf:Kanbayasi76} In this compound, itinerant Ru 4d electrons are considered to be responsible for the magnetic and transport properties.\cite{rf:Allen96}  Recent photoemission studies \cite{rf:Fujioka97,rf:Okamoto99}, however, revealed that the spectral density at the Fermi level is significantly reduced due to the strong electronic correlation, compared with that from the band calculation. Moreover, this compound shows an anomalous transport property called the ``bad metal".\cite{rf:Allen96,rf:Gunnarsson2003} The electrical resistivity increases linearly with increasing temperature, and then passes through the Ioffe-Regel limit without saturation at high temperatures. In such a high temperature region, the relation $k_{\rm F}l \sim 1$ ($l$: mean free path) holds, suggesting the breakdown of the itinerant quasi-particle picture. These experimental results indicate that both the strong correlation and the itinerant-localized duality of the Ru 4d electrons play important roles in the magnetic and transport properties.

Recently, Cao {\it et al.}\cite{rf:Cao2005} revealed in the mixed compounds SrRu$_{1-x}$Mn$_x$O$_3$ ($0\le x\le 0.6$) that the substitution of Mn for Ru suppresses the itinerant ferromagnetic phase, and then induces a new insulating phase above the critical point $x_{\rm c} = 0.39$.  For $x\ge x_{\rm c}$, a clear cusp like anomaly is observed in the temperature variations of the magnetic susceptibility, strongly suggesting the appearance of the antiferromagnetic order in the insulating phase. In SrRuO$_3$, the carrier mean free path is comparable to the lattice constant, and therefore, it is expected that even a small disorder will modify its physical properties quite appreciably. It is thus interesting to investigate the magnetic properties in the insulating phase microscopically. However, the appearance of this transition is found to be sensitive to the conditions of the sample preparation. Sahu {\it et al.}\cite{rf:Sahu2002} reported that the ferromagnetic state is still stable up to $x=0.5$, which is in sharp contrast to the results reported by Cao {\it et al.} In the present study, we first prepared the samples of SrRu$_{1-x}$Mn$_x$O$_3$ for $0.4\le x\le 0.6$ under several conditions to check the sample dependence, and then performed the powder neutron diffraction measurements.

The polycrystalline samples of SrRu$_{1-x}$Mn$_x$O$_3$ with $x=0.4$, 0.5 and 0.6 were prepared by the solid-state method with the starting materials of SrCO$_3$, RuO$_2$ and MnO.  The samples with stoichiometric compositions were first calcined at 750$^\circ$C for 4 h and shaped into pellets. They were then sintered at 1300$^\circ$C for 24 h in the ambient atmosphere. Surprisingly, X-ray powder diffraction measurements at room temperature revealed that they form the tetragonal crystal structure, in contrast to the orthorhombic one for $x\sim 0$ (Fig.\ 1). Powder patterns indicate the slight contamination of the orthorhombic phase, but it vanishes within the experimental accuracy by iterating the above sintering process tenfold. The tetragonal crystal structure observed in our samples basically coincides with the first report,\cite{rf:Xu2000} but differs from the recent ones,\cite{rf:Cao2005,rf:Sahu2002} where the orthorhombic structure observed in pure SrRuO$_3$ is kept up to $x=0.6$. We also prepared the polycrystalline samples for $x=0.5$ and 0.6 by the same heating process as that reported by Sahu {\it et al.},\cite{rf:Sahu2002} and grew a single crystal with $x\sim 0.3$ by the flux method. However, all the samples were found to form the tetragonal structure. Our preliminary investigations for the samples between $x=0$ and 0.6 indicate that the boundary between the orthorhombic phase and tetragonal phase lies between $x=0.2$ and 0.3.  The reason for the discrepancy is not clear in the present stage, probably caused by the unexpected differences in the synthesized conditions of the samples.
\begin{figure}[tbp]
\begin{center}
\includegraphics[keepaspectratio,width=0.42\textwidth]{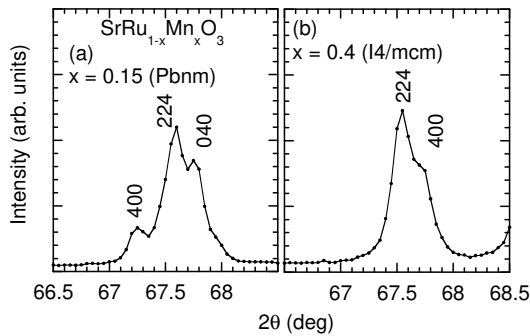}
\end{center}
\caption{X-ray diffraction profiles around the (400) and (040) peaks for SrRu$_{1-x}$Mn$_{x}$O$_3$ with (a) $x=0.15$ and (b) $x=0.4$, where the peaks are indexed by assuming the orthorhombic $Pbnm$ symmetry for (a) and the tetragonal $I4/mcm$ symmetry for (b). Note that the (040) peak does not exist independently in (b). 
}
\end{figure}

In Fig.\ 2, the temperature variations of the magnetic susceptibility $\chi (T)$ for $x=0.4$, 0.5 and 0.6, measured using a SQUID magnetometer in the temperature range between 2 K and 350 K, are shown. The hysteretic behavior is observed in $\chi(T)$ between the data under the zero-field-cooling (ZFC) and field-cooling (FC) conditions, indicating the evolution of the ferromagnetic order. The magnitude of the low-temperature magnetization is, however, estimated to be only less than 1\% of that for $x=0$. We have found that the magnitude of the hysteresis in $\chi(T)$ is much larger in the samples sintered only once, and reduced by iterating the heating process. In addition, the onset of the hysteresis is nearly independent of $x$. These properties strongly suggest that the hysteretic behavior is due to the mixture of the extrinsic ferromagnetic component with less than 1\% volume fraction. On the other hand, cusp like anomalies are observed in $\chi (T)$ at 180 K and  220 K ($\equiv T_{\rm N}$) for $x=0.5$ and 0.6 respectively, suggesting the antiferromagnetic (AF) transition. For $x=0.4$, the anomaly appears at $\sim 70\ {\rm K}$, but it is obscured by the hysteretic behavior. At high temperatures, $\chi (T)$ roughly follows the Curie-Weiss law $\chi=N\mu_{\rm eff}^2/3k_{\rm B}(T-\theta)+\chi_0$. The best fit for the $\chi (T)$ data between 280 K and 350 K yields $\mu_{\rm eff}$ values of 3.70(18) $\mu_{\rm B}$ for $x=0.4$, 3.70(29) $\mu_{\rm B}$ for $x=0.5$ and 4.12(34) $\mu_{\rm B}$ for $x=0.6$. The $\theta$ values for these samples cannot be determined for a fairly large uncertainty due to narrow fitting ranges. We have also measured the temperature variations of the electrical resistivity $\rho(T)$ for $x=0.4$, 0.5 and 0.6, and found that $\rho(T)$ shows the insulating behavior (the inset of Fig.\ 2). The properties of $\chi(T)$ and $\rho(T)$ are quite similar to those reported by Cao {\it et al.}\cite{rf:Cao2005}, suggesting that the electronic states in the samples with the tetragonal structure are nearly identical to those with the orthorhombic ones. The detailed experimental results and analyses on our samples in the entire $x$ range will be described elsewhere.\cite{rf:Kawanaka2005}
\begin{figure}[tbp]
\begin{center}
\includegraphics[keepaspectratio,width=0.45\textwidth]{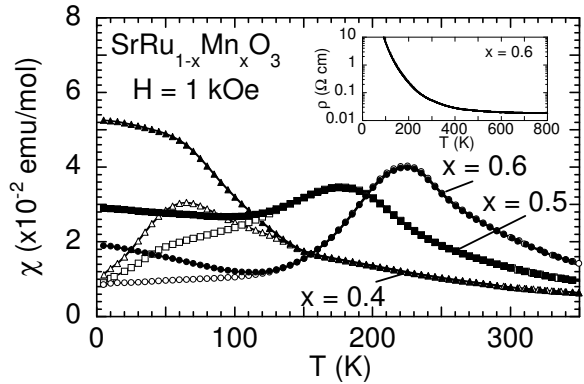}
\end{center}
\caption{Temperature variations of the magnetic susceptibility for $x=0.4$, 0.5 and 0.6. The open and solid symbols indicate the data obtained under the zero-field-cooling (ZFC) and field-cooling (FC) conditions, respectively. The inset shows the temperature variations of the electrical resistivity for $x=0.6$.
}
\end{figure}

The neutron diffraction measurements for the powdered samples with $x=0.4$, 0.5 and 0.6 were performed in the temperature range between 20 K and 290 K, using the HERMES spectrometer located at the research reactor JRR-3M of the Japan Atomic Energy Research Institute. We chose the wavelength of the incident neutron 1.82035 ${\rm \AA}$ produced by the (331) reflection of a Ge monochromator. Figure 3(a) shows the powder neutron diffraction profile for $x=0.6$ at 290 K. The diffraction patterns for $x=0.4$, 0.5 and 0.6 were analyzed using the Rietveld method with the RIETAN 2000 program.\cite{rf:Izumi2000} We assumed that the crystal structure has a tetragonal $I4/mcm$ symmetry, where ionic positions are assigned to Sr $(0,1/2, 1/4)$, Ru/Mn (0,0,0), O(1) $(0,0,1/4)$, and O(2) $(1/4+u,3/4+u,0)$. The parameters and reliability factors obtained by best fits are listed in Table I. The qualities of the refinement are slightly worse, probably due to the small mixture of the ferromagnetic phase observed in $\chi(T)$.  We have found that both the $a$ and $c$ axis lattice parameters shrink with increasing $x$, reflecting the small radius of Mn ion compared to that of Ru ion.  On the other hand, the displacement $u$ of the O(2) ion from the symmetric position decreases by increasing $x$. We note that some nuclear Bragg peaks such as the (211) and (213) reflections arise purely due to finite $u$ values. These peaks are clearly observed in the present measurements, eliminating the possibility that the samples form the tetragonal structure with the $P4/mmm$ symmetry\cite{rf:Xu2000}, where the O(2) ion occupies the position with $u=0$. 
\begin{figure}[tbp]
\begin{center}
\includegraphics[keepaspectratio,width=0.45\textwidth]{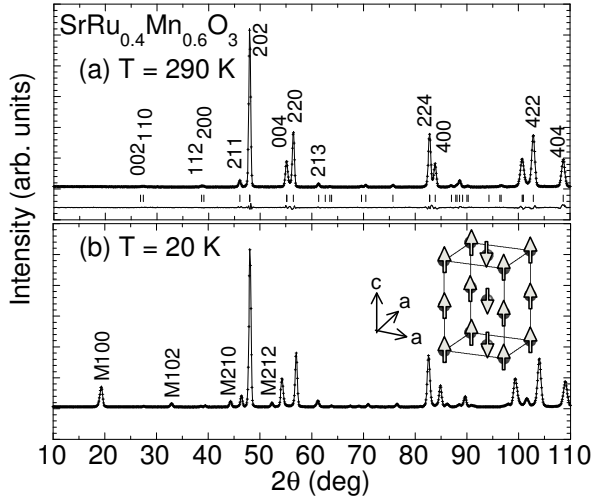}
\end{center}
\caption{Neutron powder diffraction profiles of SrRu$_{0.4}$Mn$_{0.6}$O$_3$ at (a) $T=290\ {\rm K}$ and (b) $T=20\ {\rm K}$. The mirror indices of several Bragg peaks are also shown. In (a), the calculated profile, the calculated Bragg-peak positions, and the difference between the observed and the calculated profiles are also plotted with the solid line, vertical markers, and the lower trace, respectively. The inset of (b) shows the spin arrangement on the Ru/Mn ions in the C-type AF structure, depicted in the unit cell of the tetragonal $I4/mcm$ symmetry.}
\end{figure}
\begin{table}[tbp]
\begin{center}
\begin{tabular}{cccc}
\hline
$x$ & 0.4 & 0.5 & 0.6 \\\hline
$a$ (\AA) & 5.47064(11) & 5.45744(17) & 5.44732(18)\\
$c$ (\AA) & 7.94696(23) &  7.91689(40) & 7.86858(51) \\
$u$ & 0.03200(13) &  0.02921(20) & 0.02348(25) \\
$V\ ({\rm \AA}^3)$ & 237.835(9) & 235.793(15) & 233.486(19)\\
$c/a$ & 1.4526(1) & 1.4506(2) & 1.4445(2)\\
$R_{\rm wp}(\%)$ & 3.95 & 4.93 & 4.89 \\
$R_{\rm e}(\%)$ & 2.37 & 2.63 & 2.47 \\
$S(=R_{\rm wp}/R_{\rm e})$ & 1.67 & 1.88 & 1.98\\
\hline
\end{tabular}
\end{center}
\caption{Structural parameters and reliability factors at 290 K for SrRu$_{1-x}$Mn$_x$O$_3$ with $x=0.4$, 0.5 and 0.6 obtained by the Rietveld analyses under the tetragonal $I4/mcm$ symmetry.
The parameter $u$ indicates the O(2) position $(1/4+u,3/4+u,0)$.
}
\end{table}

Figure 3(b) shows the powder neutron diffraction profile for $x=0.6$ at 20 K. We have found new Bragg reflections at positions corresponding to ($hkl$) with $h+k=odd$ and $l=even$. The peak intensities in the high scattering angle are smaller than those in the low scattering angle, strongly suggesting that these peaks originate in the magnetic order. We have not observed the other additional reflection due to a ferromagnetic or AF order within the experimental resolution. These results indicate that the system has an AF order with a modulation of $\bm{q}=(1,0,0)$, called the C-type structure in the perovskite compounds (inset of Fig.\ 3(b)). The same magnetic structure can be also derived from the powder patterns for $x=0.4$ and 0.5 at 20 K. 

It is difficult in the present study to determine exactly the polarization of the AF moment, because we have not observed the magnetic peak that completely vanishes due to the magnetic polarization factor. Since the magnetic Bragg peak intensities at the reciprocal lattice points involving the $c^*$ component are found to be rather reduced, we here assume the direction of the AF moment to be along the tetragonal $c$ axis. On this assumption, we have estimated the magnetic form factor multiplied by the magnitude of the AF moment $\mu_{\rm o}f(Q)$ ($Q=4\pi\sin\theta/\lambda$) from the integrated intensities of the magnetic Bragg reflections. The $Q$ dependences of $f(Q)$ for $x=0.4$, 0.5 and 0.6 at 20 K are plotted in Fig.\ 4. The $f(Q)$ functions show monotonic curves without an improperly large oscillation, supporting our assumption. The angle between the AF moment and the $c$ direction is estimated to be less than 5$^\circ$ even if the AF moment is canted. 
\begin{figure}[tbp]
\begin{center}
\includegraphics[keepaspectratio,width=0.45\textwidth]{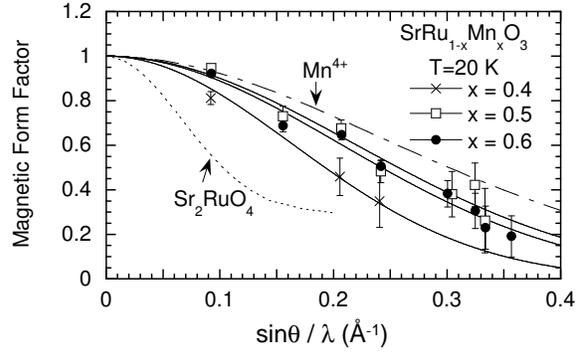}
\end{center}
\caption{Magnetic form factor at $T=20\ {\rm K}$ for SrRu$_{1-x}$Mn$_{x}$O$_3$ with $x=0.4$, 0.5 and 0.6. The experimental data for Sr$_2$RuO$_4$\cite{rf:Nagata2004} and the calculated data for Mn$^{4+}$\cite{rf:IntTable} are also shown. The solid lines are guides to the eye.}
\end{figure}

Figure 5 shows the $x$ dependence of the AF moment $\mu_{\rm o}$, the unit-cell volume $V$ and the $c/a$ ratio at 20 K. The temperature variations of $\mu_{\rm o}$ and $c/a$ for $x=0.6$ are also plotted in the insets of Fig.\ 5. The magnitude of $\mu_{\rm o}$ at 20 K for $x=0.4$ is estimated to be 1.1(2) $\mu_{\rm B}$. Interestingly, this value is found to be much smaller than $\mu_{\rm eff}$. The ratio $\mu_{\rm o}/\mu_{\rm eff}$ is estimated to be $\sim 0.34$, which is close to the value observed in pure SrRuO$_3$ ($\sim 0.4$). By further doping Mn, $\mu_{\rm o}$ increases, and reaches 2.6(1) $\mu_{\rm B}$ at $x=0.6$. At the same time, $\mu_{\rm o}/\mu_{\rm eff}$ is also enhanced to $\sim 0.63$ at $x=0.6$, suggesting that the localized characteristic of d electrons becomes strong with increasing $x$. The rather localized feature of the d electrons is also seen in the $f(Q)$ curve for $x=0.6$, which shows a slower relaxation than $f(Q)$ for the itinerant 4d electrons in the Ru$^{4+}$ state of Sr$_2$RuO$_4$\cite{rf:Nagata2004,rf:comment1} (Fig.\ 4). In the temperature scans for $x=0.6$, on the other hand, we have observed that $\mu_{\rm o}$ starts increasing at $\sim T_{\rm N}$, and shows a tendency to saturate for $T/T_{\rm N} < 0.6$. The cusp like anomaly in $\chi(T)$ is thus attributed to the AF transition. 
\begin{figure}[tbp]
\begin{center}
\includegraphics[keepaspectratio,width=0.42\textwidth]{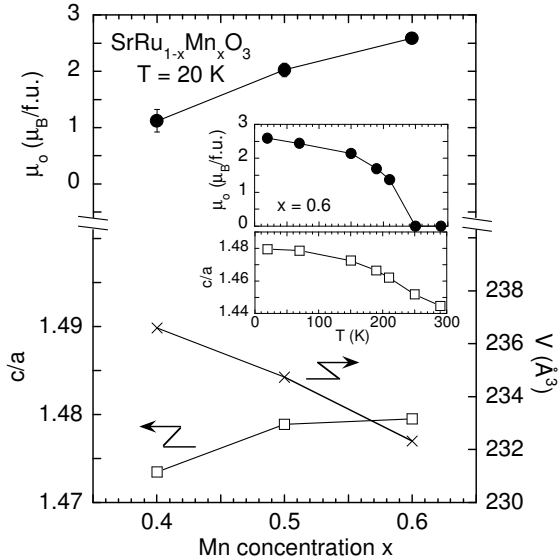}
\end{center}
\caption{$x$ variations of the AF moment $\mu_{\rm o}$, the unit-cell volume $V$ and the $c/a$ ratio at $T=20\ {\rm K}$ for SrRu$_{1-x}$Mn$_x$O$_3$. The temperature variations of $\mu_{\rm o}$ and $c/a$ ratio for $x=0.6$ are also plotted in the insets.  }
\end{figure}

We have observed that $V$ and the $c/a$ ratio at 290 K shrink with doping Mn (see Table I). $V$ at 20 K also decreases linearly with increasing $x$. The decreasing rate at 20 K ($\partial\ln V/\partial x\sim -9.0\times 10^{-2}$) is found to be the same as that at 290 K ($>T_{\rm N}$), indicating that the shrink of $V$ is not mainly caused by the evolution of $\mu_{\rm o}$. In contrast, the $c/a$ ratio at 20 K is enhanced with increasing $x$. The increasing rate $\partial\ln (c/a)/\partial x$ is roughly estimated to be $\sim 2\times 10^{-2}$. In the temperature scan for $x=0.6$, the $c/a$ ratio increases with decreasing temperature, and saturates below $\sim 70\ {\rm K}$. Since the enhancement of the $c/a$ ratio is accompanied by the development of $\mu_{\rm o}$, we find that the $c/a$ ratio is strongly coupled with the AF order.

We have found a new AF state with the C-type structure in intermediate Mn concentrations, which is completely different from the insulating AF phase with the G-type magnetic structure in pure SrMnO$_3$\cite{rf:Takeda74} as well as the itinerant ferromagnetic phase in pure SrRuO$_3$.\cite{rf:Callaghan66,rf:Kanbayasi76} Furthermore, we have observed that the ratio $\mu_{\rm o}/\mu_{\rm eff}$ is enhanced with increasing $x$ from $x=0$ to 0.6. This enhancement may be ascribed to the itinerant to localized electron transition due to the electron correlation effect, that is, the system is in the crossover region from $W/U >1$ to $W/U \ll 1$, where $W$ and $U$ are the d-band width and the on-site Coulomb potential, respectively. Under the latter condition, we expect the occurrence of AF order due to the superexchange interaction, as is argued in the Mott-type insulators. In the superexchange mechanism, the positions of the Ru/Mn and O ions play a key role in the stability of the AF order. We have found in the present neutron diffraction measurements that the intensities of some Bragg-peaks, such as the (211), (213) and (215) reflections, significantly increase with decreasing temperature. The enhancement of these peaks can be simply explained by the increase in parameter $u$, indicating that the displacement of the O(2) ion from the symmetric position increases in the AF phase. We have estimated the Ru/Mn-O(1) bond length $d_1$, the Ru/Mn-O(2) bond length $d_2$ and the Ru/Mn-O(2)-Ru/Mn bond angle $\alpha$ from the lattice parameters and the $u$ values. As temperature is lowered, $d_2$ for $x=0.6$ is reduced from 1.934 ${\rm \AA}$ (290K) to 1.921 ${\rm \AA}$ (20 K), while $d_1$ increases from 1.967 ${\rm \AA}$ to 1.996 ${\rm \AA}$. In addition, the differences in $d_1$ and $d_2$ between 20 K and 290 K increase with increasing $x$. These variations are consistent with the increase in $c/a$ ratio. From these results, we expect that the Mn doping makes the hybridization between the d electrons in Ru/Mn ions and the p electrons in O(2) ions strong through the compression in the $c$ plane. Since the condition $W/U\ll 1$ holds in the intermediate $x$ range, this effect enhances the superexchange interaction between nearest neighbor Ru/Mn ions, bringing the C-type AF order. On the other hand, $\alpha$ for $x=0.6$ slightly decreases from 169.27(11)$^{\circ}$ at 290K to 166.26(9)$^{\circ}$ at 20 K. This may also affect the magnetic structure and the magnetic excitation spectrum, although they cannot be detected in the present measurements.

It is natural to think that the valence and spin states in Ru and Mn ions are also connected with the magnetic structure. In pure SrRuO$_3$\cite{rf:Callaghan66} and SrMnO$_3$,\cite{rf:Takeda74} the realization of the low-spin Ru$^{4+}$ ($S=1$) and high-spin Mn$^{4+}$ ($S=3/2$) states was reported. By assuming a simple combination of these states, the effective moments for $x=0.4$, 0.5 and 0.6 are estimated to be 3.25 $\mu_{\rm B}$, 3.35 $\mu_{\rm B}$ and 3.46 $\mu_{\rm B}$, respectively. These values are, however, smaller than the $\mu_{\rm eff}$ values obtained from the present experiments, indicating that the other valence states or the high-spin Ru$^{4+}$ ($S=2$) state are mixed with the low-spin Ru$^{4+}$ ($S=1$) and high-spin Mn$^{4+}$ ($S=3/2$) states. We expect that such a mixing might reduce the dissimilarity of the spin states between Ru and Mn ions, and stabilize the simple colinear magnetic structure observed in SrRu$_{1-x}$Mn$_x$O$_3$ for intermediate Mn concentrations. 

One of us (M.Y.) is grateful to H.\ Amitsuka for helpful discussions and comments. This work was supported by a Grant-in-Aid for Scientific Research from the Ministry of Education, Culture, Sports, Science and Technology.



\begin{thebibliography}{99}
\bibitem{rf:comment1}We should note that the 4d electrons in Sr$_2$RuO$_4$ are known to be the high-spin state, in contrast to the low-spin state in SrRuO$_3$. In our best knowledge, there is no report on the magnetic form factor of the Ru$^{4+}$ ion with the low-spin state.
\end{thebibliography}
\end{document}